\documentclass[journal=cmatex,layout=draft]{achemso}
\setkeys{acs}{
    articletitle = false
}
\setcitestyle{square}
\usepackage{graphicx}
\usepackage[colorlinks,hyperindex]{hyperref}
\usepackage[hyperindex]{hyperref}
\hypersetup{colorlinks,citecolor=blue,linkcolor=blue,urlcolor=blue}
\usepackage{amsmath}
\usepackage[version=4]{mhchem}
\usepackage[caption=false]{subfig}
\usepackage{lipsum}
\usepackage{siunitx}
\usepackage{xcolor}
\usepackage[utf8]{inputenc}
\usepackage{mhchem}
\author{Dorothea Scheunemann}
\email{dorsche@chalmers.se}
\affiliation{Department of Chemistry and Chemical Engineering, Chalmers University of Technology, 41296 Göteborg, Sweden}
\alsoaffiliation[Linköping University]
{Complex Materials and Devices, Department of Physics, Chemistry and Biology (IFM), Linköping University, 58183 Linköping, Sweden}
\author{Vishnu Vijayakumar}
\affiliation{Universit\'{e} de Strasbourg, CNRS, ICS UPR 22, F-67000 Strasbourg, France}
\author{Huiyan Zeng}
\affiliation{Universit\'{e} de Strasbourg, CNRS, ICS UPR 22, F-67000 Strasbourg, France}
\author{Pablo Durand}
\affiliation{Universit\'{e} de Strasbourg, CNRS, ICPEES UMR 7515, F-67087 Strasbourg, France}
\author{Nicolas Leclerc}
\affiliation{Universit\'{e} de Strasbourg, CNRS, ICPEES UMR 7515, F-67087 Strasbourg, France}
\author{Martin Brinkmann}
\affiliation{Universit\'{e} de Strasbourg, CNRS, ICS UPR 22, F-67000 Strasbourg, France}
\author{Martijn Kemerink}
\affiliation{Centre for Advanced Materials, Heidelberg University, Im Neuenheimer Feld 225, 69120 Heidelberg, Germany}
\alsoaffiliation[Linköping University]
{Complex Materials and Devices, Department of Physics, Chemistry and Biology (IFM), Linköping University, 58183 Linköping, Sweden}

\title[Rubbing and drawing]
  {Rubbing and Drawing: Generic Ways to Improve the Thermoelectric Power Factor of Organic Semiconductors?}

\begin{document}


\begin{abstract}
Highly oriented polymer films can show considerable anisotropy in the thermoelectric properties leading to power factors beyond those predicted by the widely obeyed power law linking the thermopower $S$ and the electrical conductivity $\sigma$ as $S\propto\sigma^{-1/4}$. This has led to encouraging practical results with respect to the electrical conductivity, notwithstanding that the conditions necessary to enhance $\sigma$ and $S$ simultaneously are less clear. Here, kinetic Monte Carlo simulations are used to study the impact of structural anisotropy on the thermoelectric properties of disordered organic semiconductors. We find that stretching is a suitable strategy to improve the conductivity along the direction of strain, while the effect on the power factor depends on the morphology the polymer crystallizes. In general, crystalline polymers show a simultaneous increase in $\sigma$ and $S$ which is not the case for amorphous polymers. Moreover, we show that the trends resulting from simulations based on variable-range hopping are in good agreement with experiments and can describe the different functional dependencies in the $S$ versus $\sigma$ behaviour of different directions.
\end{abstract}

\maketitle

\section{Introduction}
\label{sec:intro}
\begin{figure}
\centering
\includegraphics{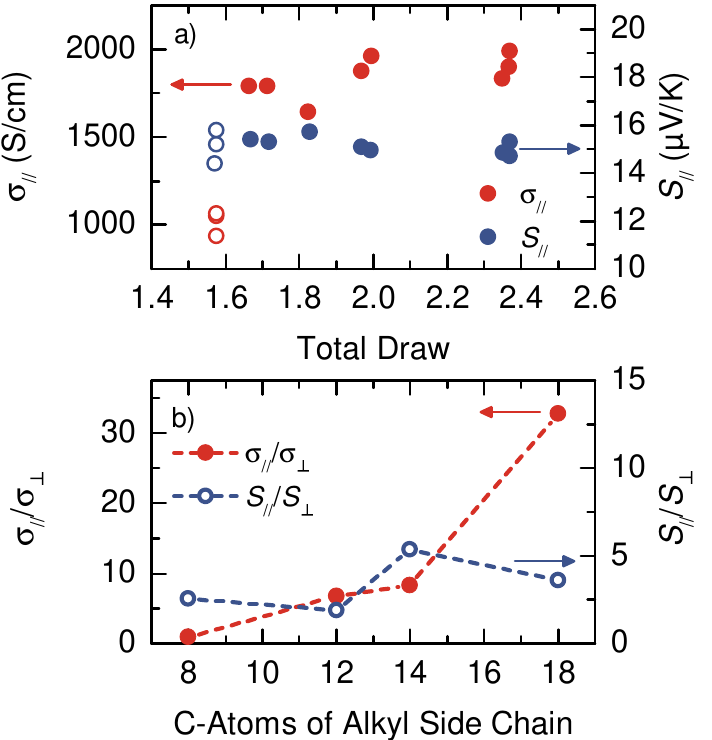}
\caption{a) Electrical conductivity (red circles) and thermopower (blue circles) of PEDOT:PSS fibers as a function of draw. PEDOT:PSS fibers were spun into 10 vol\% DMSO in IPA (open circles) and a fraction of samples was subsequently drawn through a DMSO bath (filled circles). Adapted with permission.~\cite{Sarabia-Riquelme2019} Copyright 2019, American Chemical Society b) Experimental data on PBTTT from  Ref.~\cite{Vijayakumar2019_chains} showing the anisotropy of the electrical conductivity and thermopower as a function of the alkyl side chain length at a doping time of $\SI{6}{min}$. Adapted with permission. Copyright 2019, American Chemical Society.}
\label{fig:figure01}
\end{figure}

Semiconducting organic materials have attracted increasing interest as thermoelectric (TE) converters in the recent years due to their potentially low fabrication costs and non-toxicity. As in other organic electronic devices, mastering morphology and crystallinity of polymer semiconductors is a necessity to control the performance of organic TEs. One specific case of morphology, which seems to improve both the mechanical and the electrical properties of organic TEs is the introduction of structural anisotropy. Experimentally this can be induced by orienting of the polymer backbones using high-temperature rubbing or tensile drawing. Such highly oriented conducting polymer films can show considerable increases in the electrical conductivity $\sigma$ along the direction of orientation, while the influence on the Seebeck coefficient $S$ is less clear. Experimental studies on iodine-doped PPVs \cite{Hiroshige2006} as well as on poly(3,4-ethylenedioxythiophene) polystyrene sulfonate (PEDOT:PSS) \cite{Wei2014,Sarabia-Riquelme2019} have found that the thermopower remains practically constant in the direction of orientation, as illustrated in Figure~\ref{fig:figure01}a for samples based on PEDOT:PSS fibers. In contrast, for thin films based on poly(2,5-bis(3-alkylthiophen-2-yl)thieno[3,2-\textit{b}]thiophene)(PBTTT) recent reports indicate that in-plane anisotropy can simultaneously enhance the thermopower and the electrical conductivity \cite{Vijayakumar2019, Vijayakumar2019_chains}, see Figure~\ref{fig:figure01}b, while for films of doped polythiophenes it depends on the experimental conditions whether an anisotropy in the thermopower is observed \cite{Hamidi-Sakr2017} or not \cite{Hynynen2019}.

This experimental divergence is accompanied by a lack of formal understanding. Shklovskii and Efros discussed the anisotropy of hopping conduction by applying the percolation method to the anisotropic situation in germanium subjected to a large uniaxial stress \cite{Shklovskii1984}. They argue that in this case, only one critical percolation cluster exists, which determines the transport in all directions. In consequence, the anisotropy in resistivity $\rho$ results from a pre-exponential factor and scales with $\frac{\rho_{zz}}{\rho_{xx}}=\left(\frac{a}{b}\right)^2$ where $a$ and $b$ are the decay length of an ellipsoidal-shaped wave function. The authors did not address the possibility of anisotropy in the thermopower.

Recently, Vijayakumar \textit{et al.} experimentally studied the anisotropy in the thermoelectric properties of highly oriented films based on PBTTT and P3HT, respectively \cite{Vijayakumar2019}. Based on different functional dependencies of $S(\sigma)$ in the different directions, the authors concluded that two different charge transport behaviors may govern the parallel and perpendicular directions. However, this contrasts the interpretation of Ref.~\cite{Hynynen2019} where for stretched P3HT the same transport mechanism is assumed in all directions. Hence, it is currently not evident how structural anisotropy is influencing the thermoelectric properties of disordered organic semiconductors.

Here, we combined kinetic Monte Carlo (kMC) simulations with conductivity and thermopower measurements on doped PBTTT films to rationalize the impact of anisotropy on the thermoelectric properties of disordered organic semiconductors. We find that variable-range hopping (VRH) can not only consistently describe charge and energy transport in both the parallel and the perpendicular direction but also the trends observed with polymer side chain length. Furthermore, we clarify why some materials show a simultaneous enhancement in conductivity and thermopower in the direction of orientation, while for others the Seebeck coefficient remains largely unaffected.

\section{Theoretical Framework}
\label{sec:model}
The kMC model has been extensively described before \cite{Abdalla2015,Zuo2019}. In brief, the kMC simulations within this work account for variable-range hopping on a regular or irregular (random) lattice with a mean inter-site distance of $a_\text{NN}=N_\text{0}^{-1/3}=\SI{1.8}{nm}$ with $N_\text{0}$ the total site density. Site energies are distributed according to an exponential density of states (DOS) with varying degree of disorder. An exponential form is chosen to reflect the doping-induced DOS tail that forms irrespective of the, typically Gaussian, shape of the undoped material \cite{Silver1989,Abdalla2017}. Only on-site Coulomb interactions between mobile charges are accounted for. In contrast to the approach where full Coulomb interactions (charge-charge and charge-dopant) are accounted for, as previously used to describe doping \cite{Abdalla2017, Boyle2019}, the current approach allows to keep calculation times reasonable, i.e. hours to days on a fast pc. Hopping probabilities are calculated from a Miller-Abrahams-type expression \cite{Miller1960}, while anisotropy is introduced via the localization length. For the simulations below, a standard parameter set with an attempt to hop frequency $\nu_0=\SI{E13}{s^{-1}}$, an exponential disorder of $\sigma_\text{DOS}=4\,k_\text{B}T$, an inter-site distance of $a_\text{NN}=\SI{1.8}{nm}$, a localization length of $\alpha=\alpha_0\cdot\left[ \alpha_{\text{xy}}\,\alpha_{\text{xy}}\,\alpha_{\text{z}}\right]=0.4\cdot \left[ 1\,1\,2\right] \SI{}{nm}$ a relative concentration $c=\frac{n}{N_0}=0.01$ with $n$ the absolute charge carrier concentration and $z$ being the direction of strain, a temperature $T=\SI{300}{K}$, a dielectric constant of $\varepsilon_r=2.7$  and a regular lattice is used unless indicated otherwise. 
The critical percolation network is illustrated by adding connections between sites in the order of their respective inter-site transition rates as determined from the Miller-Abrahams rates. Connections are added to the network until a set of connected bonds forms a subnetwork that spans the full simulation box in the direction in which the conductivity is probed.

\section{Results}
\begin{figure*}
\centering
\includegraphics[width=\textwidth]{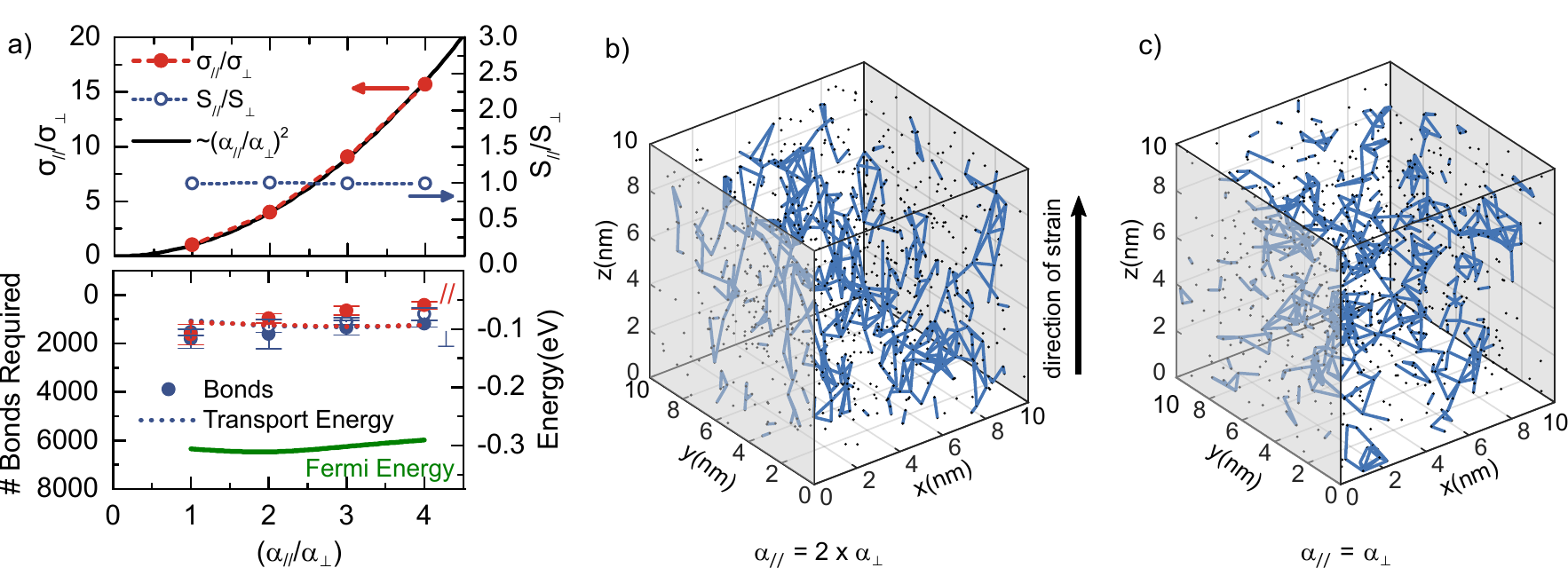}
\caption{Hopping on a random lattice at a relative carrier concentration of  $c=0.01$. a) Anisotropy in electrical conductivity $\sigma_{\parallel}/\sigma_{\perp}$ (filled red circles) and Seebeck coefficient $S_{\parallel}/S_{\perp}$ (open blue circles) (upper panel) and minimum number of bonds required until a path in the respective direction is reached (symbols) (lower panel) as a function of the anisotropy ratio. The dotted lines in the lower panel present the transport energy $E_\text{tr}$ for the parallel (red) and perpendicular (blue) direction, respectively. The green line represents the Fermi energy $E_\text{F}$. b, c) Critical percolation cluster (blue lines) for hopping on a random lattice with $\alpha=0.4\cdot\left[ 1\,1\,2\right] \SI{}{nm}$ (b) and $\alpha=0.4\cdot\left[ 1\,1\,1\right] \SI{}{nm}$ (c). Black dots represent available sites, gray surfaces indicate the $yz$-planes.}
\label{fig:figure02}
\end{figure*}

In polymer films, the crystallization behavior, which has a major impact on nanomorphology, is often a key issue with respect to performance control. Therefore, we discuss first the influence of structural differences in terms of different lattice structures. The upper panel of Figure~\ref{fig:figure02}a shows the anisotropy of conductivity and thermopower as a function of the anisotropy ratio $\alpha_{\parallel}/\alpha_{\perp}$ for a random lattice. We find that this type of lattice results in a thermopower which is independent on the direction of transport, while the anisotropy in conductivity scales quadratic with the anisotropy ratio, as predicted by Efros und Shklovskii \cite{Shklovskii1984}. In agreement with their assumption, the critical percolation network for this case is found to be similar in all directions of transport as depicted in Figure~\ref{fig:figure02}b and also similar to the isotropic case (see panel c). This is also reflected in the percolation threshold, here defined as the minimum number of bonds required until a path in the respective direction is formed. As shown in the lower panel of Figure~\ref{fig:figure02}a this number is independent of the direction of transport, which correlates with the behavior of the transport energy and a thermopower that is independent on the direction of transport.

\begin{figure*}
\centering
\includegraphics[width=\textwidth]{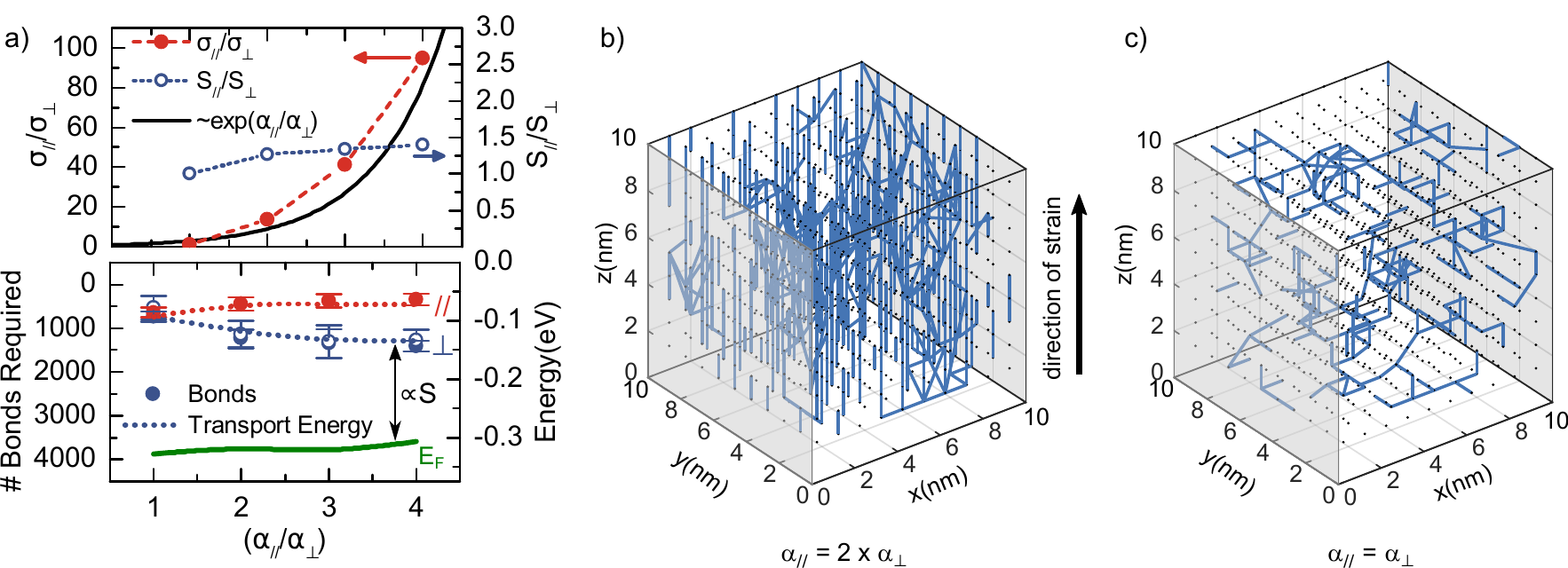}
\caption{Hopping on a regular lattice at a relative carrier concentration of  $c=0.01$. a) Anisotropy in electrical conductivity $\sigma_{\parallel}/\sigma_{\perp}$ (filled red circles) and Seebeck coefficient $S_{\parallel}/S_{\perp}$ (open blue circles) (upper panel) and minimum number of bonds required until a path in the respective direction is reached (symbols) (lower panel) as a function of the anisotropy ratio. The dotted lines in the lower panel present the transport energy $E_\text{tr}$ for the parallel (red) and perpendicular (blue) direction, respectively. The green line represents the Fermi energy $E_\text{F}$. b, c) Critical percolation cluster (blue lines) for hopping on a regular lattice with $\alpha=0.4\cdot\left[ 1\,1\,2\right] \SI{}{nm}$ (b) and $\alpha=0.4\cdot\left[ 1\,1\,1\right] \SI{}{nm}$ (c). Black dots represent available sites, gray surfaces indicate the $yz$-planes.}
\label{fig:figure03}
\end{figure*}

In contrast, for a regular lattice the critical percolation network has a distinct orientation in the direction of strain, see Figure~\ref{fig:figure03}b and clearly differs from the percolation network for the isotropic case, see panel c. This results in an anisotropy in the thermopower which is increasing with anisotropy ratio and an anisotropy in conductivity that scales roughly exponentially with the anisotropy ratio, as depicted in Figure~\ref{fig:figure03}a. Qualitatively, the observed behavior can be understood as follows. The thermopower is given by $S\propto (E_\text{F}-E_\text{tr})/T$ with $E_\text{F}$ the Fermi energy, $E_\text{tr}$ the transport energy, and $T$ the temperature and therefore a direct measure of the energy difference $\Delta E=E_\text{F}-E_\text{tr}$. For a highly anisotropic localization radius that is small compared to the nearest neighbor distance on a regular lattice, carriers are ‘forced’ to hop along the parallel direction and therefore have a reduced ability to optimize their path with respect to energy. This leads to the highly anisotropic percolating network as depicted in Figure~\ref{fig:figure03}b, and in a percolation threshold that is reduced in the parallel compared to the perpendicular direction and concomitantly an upwards shift of the transport energy in this direction, as depicted in the lower panel of Figure~\ref{fig:figure03}a. Finally, this results in an increase in $S_{\parallel}$ compared to $S_{\perp}$. In consequence, the introduction of structural anisotropy enhances the power factor $PF=\sigma S^2$ in the parallel direction, as also observed experimentally for aligned PBTTT and P3HT films\cite{Vijayakumar2019_chains,Vijayakumar2019,Hamidi-Sakr2017}. As depicted in Figure S1 in the Supplemental Material, this enhancement is especially pronounced for the case of a regular lattice.

The investigations above clarify the diverging results observed experimentally for the anisotropy of the thermopower. For more amorphous polymers like PEDOT:PSS, a random lattice can reasonably be assumed \cite{Nardes2007,Ouyang2015}. In the simulations above, this assumption adequately reproduces the experimental finding of an isotropic thermopower shown in Ref.~\cite{Wei2014,Sarabia-Riquelme2019}but also the correlation of hopping length in parallel and perpendicular direction to conductivity in these directions found by Nardes \textit{et al.} \cite{Nardes2007}. 
In contrast, highly crystalline polymers like PBTTT show strong spatial correlations and therefore are not random systems. Hence, the anisotropy in thermopower must, in first order, be described using a regular lattice. Whether semicrystalline polymers like P3HT can be classified to one or the other group will likely depend on the preparation conditions and the corresponding degree of crystallinity.

\begin{figure}
\centering
\includegraphics{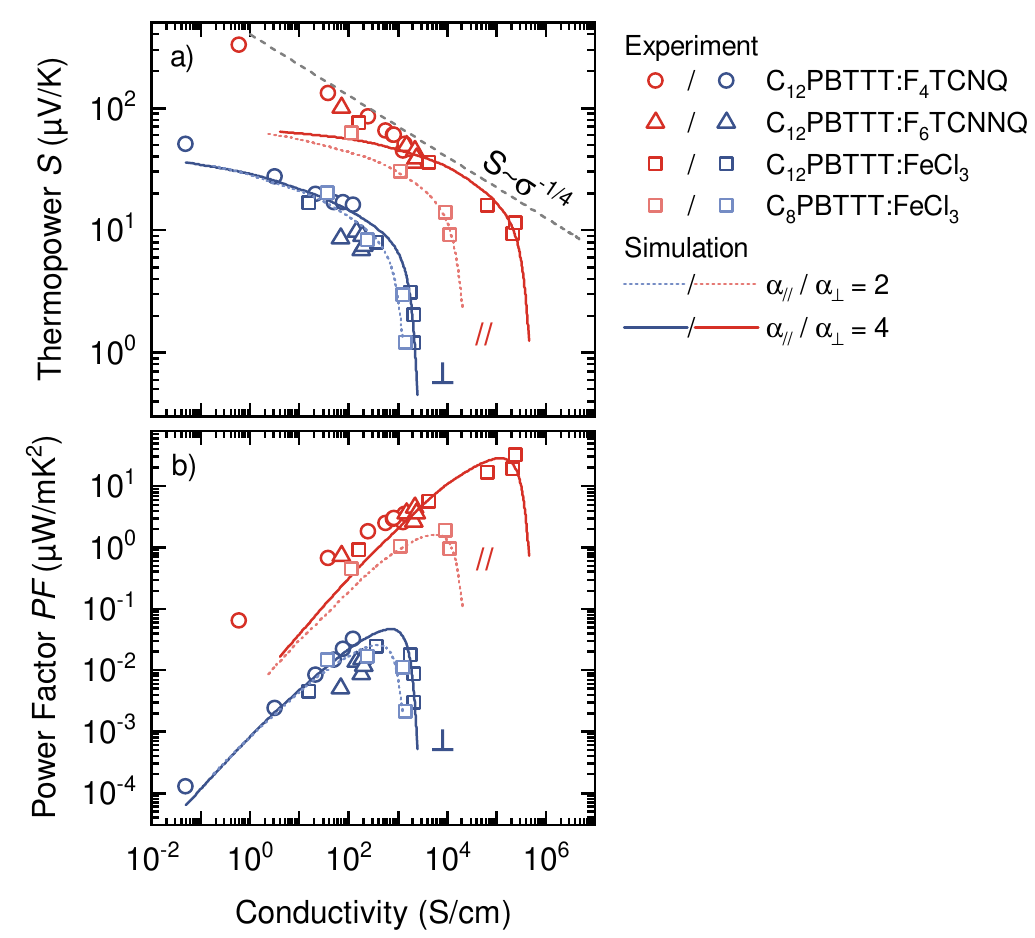}
\caption{a) Thermopower $S$ and b) power factor $PF$ versus conductivity of highly oriented thin films based on C$_{8}$PBTTT doped with FeCl$_{3}$ (bright squares) and C$_{12}$PBTTT doped with F$_{4}$TCNQ (circles), F$_{6}$TCNQ (triangles) and FeCl$_{3}$ (squares). Lines correspond to simulations calculated from the kMC model applied to a regular lattice, with an attempt to hop frequency of $\nu_0=\SI{3E14}{s^{-1}}$ and a thermopower rescaled by a factor of 0.05 (0.075) in the perpendicular (parallel) direction. Dotted lines represent an anisotropy ratio of $\alpha_{\parallel}/\alpha_{\perp}=2$, while full lines show the result for an anisotropy ratio of $\alpha_{\parallel}/\alpha_{\perp}=4$. $\parallel$ and $\perp$ refer to the parallel and perpendicular direction, respectively. The dashed gray line shows the $-1/4$ power-law relationship observed experimentally.}
\label{fig:figure04}
\end{figure}

To test the predictions of the model on real devices we used thin films based on a series of doped PBTTT polymers that differ by the length of their alkyl side chains, from n-octyl to n-dodecyl (C$_{8}$, C$_{12}$). A short summary of the experiments is given in the methods section, for full details we refer to Refs.~\cite{Hamidi-Sakr2017,Koech2010,Vijayakumar2020}. Figure~\ref{fig:figure04} shows the experimentally determined conductivity dependence of the thermopower and power factor together with the results of the kMC simulations. In the perpendicular direction, the experimentally determined thermopower follow a universal curve, independent of the polymer or the dopant used. This is not the case for the parallel direction. Here, the experimental data follow at lower conductivities the  “universal” $-1/4$ power-law relationship that was previously noted by Glaudell \textit{et al.} \cite{Glaudell2015} and explained by Abdalla \textit{et al.} \cite{Abdalla2017} but show a roll-off that depends on the length of the alkyl side chain. Both the trend in the perpendicular and the parallel direction can be well reproduced by our kMC model (lines), when assuming that the anisotropy ratio is varied from $\alpha_{\parallel}/\alpha_{\perp}=2$ (dotted lines) to $\alpha_{\parallel}/\alpha_{\perp}=4$ (full lines). This increase in the anisotropy ratio $\alpha_{\parallel}/\alpha_{\perp}$, enhances the anisotropy in conductivity and thermopower, in accordance with the results shown in Figure~\ref{fig:figure02}b. The assumption of a variation in anisotropy ratio with alkyl side chain length of the polymer is reasonable, when considering that the rubbed films consist mainly of aligned face-on oriented crystals and the interlayer spacing increases with alkyl side chain length \cite{Vijayakumar2019_chains}. Detailed investigations on how the polymer structure relates to the anisotropy factor will be the topic of further research. We attribute the fact that the experimentally observed $-1/4$ power-law relationship is not well reproduced by the kMC simulations to the approximate treatment of the Coulomb interaction, as discussed in the Theoretical Framework section; in Ref.~\cite{Boyle2019} it is shown that a more realistic DOS shape in an otherwise similar model does reproduce the $-1/4$ power-law relationship. 

To scale the absolute values, the attempt to hop frequency $\nu_0$ was set to a value of $\nu_0=\SI{3E14}{s^{-1}}$, while the thermopower was rescaled by a factor of 0.05 (0.075) in the perpendicular (parallel) direction. The latter indicates that the kMC model is not perfectly adjusted on an absolute energy scale, implying that disorder is most likely overestimated. A more quantitative description will be in the focus of upcoming work. However, the model is able to capture the qualitative behavior of the data, without the need to assume different transport mechanisms in the in- and out-of-plane directions. 

\begin{figure}
\centering
\includegraphics{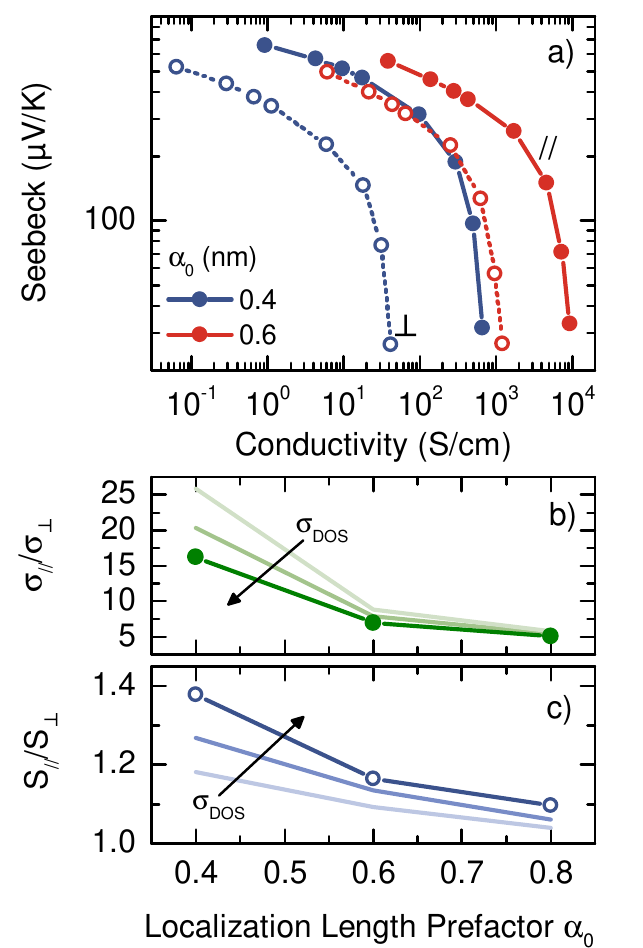}%
\caption{a) Simulated thermopower $S$ as a function of the electrical conductivity $\sigma$ for a regular lattice with different localization length prefactors $\alpha_0$. $\parallel$ (filled circles) and $\perp$ (open circles) refer to the parallel and perpendicular directions, respectively. b, c) Anisotropy in electrical conductivity $\sigma_{\parallel}/\sigma_{\perp}$ (b) and Seebeck coefficient $S_{\parallel}/S_{\perp}$ (c) in the parallel and perpendicular direction as a function of the localization length prefactor $\alpha_0$ and for a variation of the energetic disorder $\sigma_\text{DOS}$, ranging from $\sigma_\text{DOS}=2\,k_\text{B}T$ to $4\,k_\text{B}T$, $\alpha = \alpha_0 \cdot[1 1 2]$.}
\label{fig:figure05}
\end{figure}

Figure~\ref{fig:figure05} shows the anisotropy in the $S$ versus $\sigma$ relation with respect to the localization length prefactor $\alpha_0$. Increasing the overall localization length results in an increase in conductivity as could be expected for a stronger wavefunction overlap between adjacent sides. At the same time anisotropy in both $\sigma$ and $S$ is reduced with increasing $\alpha_0$ as illustrated in panel b and c. Once the actual tunneling distances, that scale with $\alpha$, get larger than the typical inter-site distance, the effect of the correlations in the (regular) lattice get washed out. Equivalently, the effect of increasing $\alpha_0$ may be understood as making the system behave more like a true VRH system with significant hopping to non-nearest sites for which only a mild anisotropy in conductivity, but not in thermopower, is observed, c.f. Figure~\ref{fig:figure02}. The exact value of the anisotropy depends moreover on the energetic disorder $\sigma_\text{DOS}$. Specifically, increasing the energetic disorder decreases the anisotropy in the conductivity (see panel b) but at the same time increases the anisotropy in the thermopower, see panel c. However, with respect to high power factors $PF=S^2\sigma$ a low energetic disorder is preferential as this leads to an increase in the absolute value of the conductivity that is higher than the decrease in $S^2$.

\section{Conclusion}
With the help of kinetic Monte Carlo simulations we systematically investigated the impact of structural anisotropy in disordered organic semiconductors. By analyzing the impact of the parameters affecting the length scales and the structural order of such systems we examine under which conditions creating structural anisotropy by e.g. rubbing or drawing, is a suitable strategy to enhance the power factor of organic thermoelectric materials. 

Even though the quantitative relation between ‘stretching’ and the absolute values of the localization length is not clear, we could show here that qualitative agreement with key experimental observations can be obtained by introducing structural anisotropy via an increase of the localization length in the parallel direction and keeping $\alpha_{\perp}$ constant. In contrast, an approach that keeps the sum or the product of  $\alpha_{\perp}$ and $\alpha_{\parallel}$ constant leads to a decrease in $\sigma_{\perp}$, which is not observed in experiments.

Furthermore, we showed that well known strategies such as aiming for low energetic disorder and high localization length which lead in general to high power factors, are also the method of choice in the case of structural anisotropy. However, structural anisotropy offers an additional tool. On the one hand, the power factor (in parallel direction) can be further increased compared to the isotropic case as the commonly observed trade-off between conductivity increase and thermopower decrease can be avoided, and power factors beyond values predicted by the empirical power law $S\propto\sigma^{-1/4}$ can be achieved. Particularly promising in this context are highly crystalline polymers, corresponding to a regular hopping lattice in our model, that allow a simultaneous, anisotropy-induced increase in the thermopower and conductivity and therefore the highest power factors. On the other hand, the use of structural anisotropy offers an alternative to increasing the overall localization length, since when using a highly crystalline material, increasing $\alpha$ in one direction by stretching leads to similarly high power factors in that direction as increasing the overall $\alpha$ would do. Technically, this can be advantageous, since methods to create structural anisotropy in polymer films are well known\cite{Biniek2014,Brinkmann2014,Pandey2019}, while it may be more difficult to increase the overall localization length.

\section{Experimental Section}
\label{sec:exp}
C$_8$PBTTT (M$_\text{w}$=\SI{24000}{g/mol}, M$_\text{n}$=\SI{13000}{g/mol}) and C$_{12}$PBTTT (M$_\text{w}$=\SI{45000}{g/mol}, M$_\text{n}$=\SI{26000}{g/mol}) were synthesized according to the synthetic path given in references \cite{Vijayakumar2019} and  \citep{Vijayakumar2019_chains}. The F$_6$TCNNQ dopant was synthesized following reference \citep{Koech2010}. High temperature rubbing of PBTTT films was performed using the protocol defined in references \cite{Hamidi-Sakr2017} and \citep{Biniek2014}. 30-\SI{60}{nm}-thick polymer films were rubbed at \SI{125}{\celsius} on a sacrificial layer of NaPSS on clean glass substrates. The alignment was quantified by polarized UV-vis-NIR spectroscopy using a Cary 5000 spectrometer. The aligned PBTTT films were floated on distilled water and recovered on pre-patterned glass substrates for four-point conductivity and Seebeck coefficient measurements.\cite{Hamidi-Sakr2017} The deposition of gold contacts and the geometry of the electrode patterns are described in reference \cite{Hamidi-Sakr2017}. Sequential doping was performed in a Jacomex glovebox ($p_{\ce{O2}}< \SI{1}{ppm}$ and $p_{\ce{H2O}}<\SI{1}{ppm}$) using anhydrous acetonitrile (F$_4$TCNQ and F$_6$TCNNQ) and nitromethane (FeCl$_3$) following the protocol described in reference \cite{Vijayakumar2019}. All solvents were used as received from Sigma Aldrich. Charge conductivity and Seebeck coefficients were probed using a Keithley 4200 source meter and a Lab Assistant Semi-Probe station along and perpendicular to the rubbing directions just after doping (to avoid aging of the samples).\cite{Vijayakumar2019_chains,Vijayakumar2019} Additional experimental details can be found in a forthcoming publication\cite{Vijayakumar2020}

\section{Supporting Information }
Supporting Information is available from the Wiley Online Library or from the author.

\begin{acknowledgement}
D.~S. acknowledges funding through the European Union’s Horizon 2020 research and innovation program under the Marie Sk\l{}odowska-Curie grant agreement No 799477.  Financial supports from the ANR Anisotherm (ANR-17-CE05-0012) and CNRS grant PEPS Thermobody are gratefully acknowledged.
\end{acknowledgement}

\bibliography{references}

\providecommand{\latin}[1]{#1}
\makeatletter
\providecommand{\doi}
  {\begingroup\let\do\@makeother\dospecials
  \catcode`\{=1 \catcode`\}=2 \doi@aux}
\providecommand{\doi@aux}[1]{\endgroup\texttt{#1}}
\makeatother
\providecommand*\mcitethebibliography{\thebibliography}
\csname @ifundefined\endcsname{endmcitethebibliography}
  {\let\endmcitethebibliography\endthebibliography}{}
\begin{mcitethebibliography}{23}
\providecommand*\natexlab[1]{#1}
\providecommand*\mciteSetBstSublistMode[1]{}
\providecommand*\mciteSetBstMaxWidthForm[2]{}
\providecommand*\mciteBstWouldAddEndPuncttrue
  {\def\EndOfBibitem{\unskip.}}
\providecommand*\mciteBstWouldAddEndPunctfalse
  {\let\EndOfBibitem\relax}
\providecommand*\mciteSetBstMidEndSepPunct[3]{}
\providecommand*\mciteSetBstSublistLabelBeginEnd[3]{}
\providecommand*\EndOfBibitem{}
\mciteSetBstSublistMode{f}
\mciteSetBstMaxWidthForm{subitem}{(\alph{mcitesubitemcount})}
\mciteSetBstSublistLabelBeginEnd
  {\mcitemaxwidthsubitemform\space}
  {\relax}
  {\relax}

\bibitem[Hiroshige \latin{et~al.}(2006)Hiroshige, Ookawa, and
  Toshima]{Hiroshige2006}
Hiroshige,~Y.; Ookawa,~M.; Toshima,~N. \emph{Synth. Met.} \textbf{2006},
  \emph{156}, 1341--1347\relax
\mciteBstWouldAddEndPuncttrue
\mciteSetBstMidEndSepPunct{\mcitedefaultmidpunct}
{\mcitedefaultendpunct}{\mcitedefaultseppunct}\relax
\EndOfBibitem
\bibitem[Wei \latin{et~al.}(2014)Wei, Mukaida, Kirihara, and Ishida]{Wei2014}
Wei,~Q.; Mukaida,~M.; Kirihara,~K.; Ishida,~T. \emph{ACS Macro Letters}
  \textbf{2014}, \emph{3}, 948--952\relax
\mciteBstWouldAddEndPuncttrue
\mciteSetBstMidEndSepPunct{\mcitedefaultmidpunct}
{\mcitedefaultendpunct}{\mcitedefaultseppunct}\relax
\EndOfBibitem
\bibitem[Sarabia-Riquelme \latin{et~al.}(2019)Sarabia-Riquelme, Shahi, Brill,
  and Weisenberger]{Sarabia-Riquelme2019}
Sarabia-Riquelme,~R.; Shahi,~M.; Brill,~J.~W.; Weisenberger,~M.~C. \emph{ACS
  Appl. Polym. Mater.} \textbf{2019}, \emph{1}, 2157--2167\relax
\mciteBstWouldAddEndPuncttrue
\mciteSetBstMidEndSepPunct{\mcitedefaultmidpunct}
{\mcitedefaultendpunct}{\mcitedefaultseppunct}\relax
\EndOfBibitem
\bibitem[Vijayakumar \latin{et~al.}(2019)Vijayakumar, Zhong, Untilova, Bahri,
  Herrmann, Biniek, Leclerc, and Brinkmann]{Vijayakumar2019}
Vijayakumar,~V.; Zhong,~Y.; Untilova,~V.; Bahri,~M.; Herrmann,~L.; Biniek,~L.;
  Leclerc,~N.; Brinkmann,~M. \emph{Adv. Energy Mater.} \textbf{2019}, \emph{9},
  1900266\relax
\mciteBstWouldAddEndPuncttrue
\mciteSetBstMidEndSepPunct{\mcitedefaultmidpunct}
{\mcitedefaultendpunct}{\mcitedefaultseppunct}\relax
\EndOfBibitem
\bibitem[Vijayakumar \latin{et~al.}(2019)Vijayakumar, Zaborova, Biniek, Zeng,
  Herrmann, Carvalho, Boyron, Leclerc, and Brinkmann]{Vijayakumar2019_chains}
Vijayakumar,~V.; Zaborova,~E.; Biniek,~L.; Zeng,~H.; Herrmann,~L.;
  Carvalho,~A.; Boyron,~O.; Leclerc,~N.; Brinkmann,~M. \emph{ACS Appl. Mater.
  Interfaces} \textbf{2019}, \emph{11}, 4942--4953\relax
\mciteBstWouldAddEndPuncttrue
\mciteSetBstMidEndSepPunct{\mcitedefaultmidpunct}
{\mcitedefaultendpunct}{\mcitedefaultseppunct}\relax
\EndOfBibitem
\bibitem[Hamidi-Sakr \latin{et~al.}(2017)Hamidi-Sakr, Biniek, Bantignies,
  Maurin, Herrmann, Leclerc, Lévêque, Vijayakumar, Zimmermann, and
  Brinkmann]{Hamidi-Sakr2017}
Hamidi-Sakr,~A.; Biniek,~L.; Bantignies,~J.-L.; Maurin,~D.; Herrmann,~L.;
  Leclerc,~N.; Lévêque,~P.; Vijayakumar,~V.; Zimmermann,~N.; Brinkmann,~M.
  \emph{Adv. Funct. Mater.} \textbf{2017}, \emph{27}, 1700173\relax
\mciteBstWouldAddEndPuncttrue
\mciteSetBstMidEndSepPunct{\mcitedefaultmidpunct}
{\mcitedefaultendpunct}{\mcitedefaultseppunct}\relax
\EndOfBibitem
\bibitem[Hynynen \latin{et~al.}(2019)Hynynen, Järsvall, Kroon, Zhang, Barlow,
  Marder, Kemerink, Lund, and Müller]{Hynynen2019}
Hynynen,~J.; Järsvall,~E.; Kroon,~R.; Zhang,~Y.; Barlow,~S.; Marder,~S.~R.;
  Kemerink,~M.; Lund,~A.; Müller,~C. \emph{ACS Macro Lett.} \textbf{2019},
  \emph{8}, 70--76\relax
\mciteBstWouldAddEndPuncttrue
\mciteSetBstMidEndSepPunct{\mcitedefaultmidpunct}
{\mcitedefaultendpunct}{\mcitedefaultseppunct}\relax
\EndOfBibitem
\bibitem[Shklovskii and Efros(1984)Shklovskii, and Efros]{Shklovskii1984}
Shklovskii,~B.~I.; Efros,~A.~L. \emph{Electronic Properties of Doped
  Semiconductors}; Springer-Verlag Berlin, 1984\relax
\mciteBstWouldAddEndPuncttrue
\mciteSetBstMidEndSepPunct{\mcitedefaultmidpunct}
{\mcitedefaultendpunct}{\mcitedefaultseppunct}\relax
\EndOfBibitem
\bibitem[Abdalla \latin{et~al.}(2015)Abdalla, van~de Ruit, and
  Kemerink]{Abdalla2015}
Abdalla,~H.; van~de Ruit,~K.; Kemerink,~M. \emph{Sci. Rep.} \textbf{2015},
  \emph{5}, 16870\relax
\mciteBstWouldAddEndPuncttrue
\mciteSetBstMidEndSepPunct{\mcitedefaultmidpunct}
{\mcitedefaultendpunct}{\mcitedefaultseppunct}\relax
\EndOfBibitem
\bibitem[Zuo \latin{et~al.}(2019)Zuo, Abdalla, and Kemerink]{Zuo2019}
Zuo,~G.; Abdalla,~H.; Kemerink,~M. \emph{Adv. Electron. Mater.} \textbf{2019},
  \emph{0}, 1800821\relax
\mciteBstWouldAddEndPuncttrue
\mciteSetBstMidEndSepPunct{\mcitedefaultmidpunct}
{\mcitedefaultendpunct}{\mcitedefaultseppunct}\relax
\EndOfBibitem
\bibitem[Silver \latin{et~al.}(1989)Silver, Pautmeier, and
  B\"assler]{Silver1989}
Silver,~M.; Pautmeier,~L.; B\"assler,~H. \emph{Solid State Commun.}
  \textbf{1989}, \emph{72}, 177--180\relax
\mciteBstWouldAddEndPuncttrue
\mciteSetBstMidEndSepPunct{\mcitedefaultmidpunct}
{\mcitedefaultendpunct}{\mcitedefaultseppunct}\relax
\EndOfBibitem
\bibitem[Abdalla \latin{et~al.}(2017)Abdalla, Zuo, and Kemerink]{Abdalla2017}
Abdalla,~H.; Zuo,~G.; Kemerink,~M. \emph{Phys. Rev. B} \textbf{2017},
  \emph{96}, 241202\relax
\mciteBstWouldAddEndPuncttrue
\mciteSetBstMidEndSepPunct{\mcitedefaultmidpunct}
{\mcitedefaultendpunct}{\mcitedefaultseppunct}\relax
\EndOfBibitem
\bibitem[Boyle \latin{et~al.}(2019)Boyle, Upadhyaya, Wang, Renna, Lu-D\'{i}az,
  Pyo~Jeong, Hight-Huf, Korugic-Karasz, Barnes, Aksamija, and
  Venkataraman]{Boyle2019}
Boyle,~C.~J.; Upadhyaya,~M.; Wang,~P.; Renna,~L.~A.; Lu-D\'{i}az,~M.;
  Pyo~Jeong,~S.; Hight-Huf,~N.; Korugic-Karasz,~L.; Barnes,~M.~D.;
  Aksamija,~Z.; Venkataraman,~D. \emph{Nat. Commun.} \textbf{2019}, \emph{10},
  2827\relax
\mciteBstWouldAddEndPuncttrue
\mciteSetBstMidEndSepPunct{\mcitedefaultmidpunct}
{\mcitedefaultendpunct}{\mcitedefaultseppunct}\relax
\EndOfBibitem
\bibitem[Miller and Abrahams(1960)Miller, and Abrahams]{Miller1960}
Miller,~A.; Abrahams,~E. \emph{Phys. Rev.} \textbf{1960}, \emph{120},
  745--755\relax
\mciteBstWouldAddEndPuncttrue
\mciteSetBstMidEndSepPunct{\mcitedefaultmidpunct}
{\mcitedefaultendpunct}{\mcitedefaultseppunct}\relax
\EndOfBibitem
\bibitem[Nardes \latin{et~al.}(2007)Nardes, Kemerink, and Janssen]{Nardes2007}
Nardes,~A.~M.; Kemerink,~M.; Janssen,~R.~A.~J. \emph{Phys. Rev. B}
  \textbf{2007}, \emph{76}, 085208\relax
\mciteBstWouldAddEndPuncttrue
\mciteSetBstMidEndSepPunct{\mcitedefaultmidpunct}
{\mcitedefaultendpunct}{\mcitedefaultseppunct}\relax
\EndOfBibitem
\bibitem[Ouyang \latin{et~al.}(2015)Ouyang, Musumeci, Jafari, Ederth, and
  Ingan\"as]{Ouyang2015}
Ouyang,~L.; Musumeci,~C.; Jafari,~M.~J.; Ederth,~T.; Ingan\"as,~O. \emph{ACS
  Appl. Mater. Interfaces} \textbf{2015}, \emph{7}, 19764--19773\relax
\mciteBstWouldAddEndPuncttrue
\mciteSetBstMidEndSepPunct{\mcitedefaultmidpunct}
{\mcitedefaultendpunct}{\mcitedefaultseppunct}\relax
\EndOfBibitem
\bibitem[Koech \latin{et~al.}(2010)Koech, Padmaperuma, Wang, Swensen,
  Polikarpov, Darsell, Rainbolt, and Gaspar]{Koech2010}
Koech,~P.~K.; Padmaperuma,~A.~B.; Wang,~L.; Swensen,~J.~S.; Polikarpov,~E.;
  Darsell,~J.~T.; Rainbolt,~J.~E.; Gaspar,~D.~J. \emph{Chem. Mater.}
  \textbf{2010}, \emph{22}, 3926--3932\relax
\mciteBstWouldAddEndPuncttrue
\mciteSetBstMidEndSepPunct{\mcitedefaultmidpunct}
{\mcitedefaultendpunct}{\mcitedefaultseppunct}\relax
\EndOfBibitem
\bibitem[Vijayakumar \latin{et~al.}()Vijayakumar, Durand, Zeng, Untilova,
  Herrmann, Algayer, Leclerc, and Brinkmann]{Vijayakumar2020}
Vijayakumar,~V.; Durand,~P.; Zeng,~H.; Untilova,~V.; Herrmann,~L.; Algayer,~P.;
  Leclerc,~N.; Brinkmann,~M. \emph{manuscript in preparation.} \relax
\mciteBstWouldAddEndPunctfalse
\mciteSetBstMidEndSepPunct{\mcitedefaultmidpunct}
{}{\mcitedefaultseppunct}\relax
\EndOfBibitem
\bibitem[Glaudell \latin{et~al.}(2015)Glaudell, Cochran, Patel, and
  Chabinyc]{Glaudell2015}
Glaudell,~A.~M.; Cochran,~J.~E.; Patel,~S.~N.; Chabinyc,~M.~L. \emph{Adv.
  Energy Mater.} \textbf{2015}, \emph{5}, 1401072\relax
\mciteBstWouldAddEndPuncttrue
\mciteSetBstMidEndSepPunct{\mcitedefaultmidpunct}
{\mcitedefaultendpunct}{\mcitedefaultseppunct}\relax
\EndOfBibitem
\bibitem[Biniek \latin{et~al.}(2014)Biniek, Pouget, Djurado, Gonthier, Tremel,
  Kayunkid, Zaborova, Crespo-Monteiro, Boyron, Leclerc, Ludwigs, and
  Brinkmann]{Biniek2014}
Biniek,~L.; Pouget,~S.; Djurado,~D.; Gonthier,~E.; Tremel,~K.; Kayunkid,~N.;
  Zaborova,~E.; Crespo-Monteiro,~N.; Boyron,~O.; Leclerc,~N.; Ludwigs,~S.;
  Brinkmann,~M. \emph{Macromolecules} \textbf{2014}, \emph{47},
  3871--3879\relax
\mciteBstWouldAddEndPuncttrue
\mciteSetBstMidEndSepPunct{\mcitedefaultmidpunct}
{\mcitedefaultendpunct}{\mcitedefaultseppunct}\relax
\EndOfBibitem
\bibitem[Brinkmann \latin{et~al.}(2014)Brinkmann, Hartmann, Biniek, Tremel, and
  Kayunkid]{Brinkmann2014}
Brinkmann,~M.; Hartmann,~L.; Biniek,~L.; Tremel,~K.; Kayunkid,~N.
  \emph{Macromol. Rapid Commun.} \textbf{2014}, \emph{35}, 9--26\relax
\mciteBstWouldAddEndPuncttrue
\mciteSetBstMidEndSepPunct{\mcitedefaultmidpunct}
{\mcitedefaultendpunct}{\mcitedefaultseppunct}\relax
\EndOfBibitem
\bibitem[Pandey \latin{et~al.}(2019)Pandey, Kumari, Nagamatsu, and
  Pandey]{Pandey2019}
Pandey,~M.; Kumari,~N.; Nagamatsu,~S.; Pandey,~S.~S. \emph{J. Mater. Chem. C}
  \textbf{2019}, \emph{7}, 13323--13351\relax
\mciteBstWouldAddEndPuncttrue
\mciteSetBstMidEndSepPunct{\mcitedefaultmidpunct}
{\mcitedefaultendpunct}{\mcitedefaultseppunct}\relax
\EndOfBibitem
\end{mcitethebibliography}

\newpage

\newpage
\noindent
\textbf{Table of Content Entry:} Kinetic Monte Carlo simulations are used to elucidate how structural anisotropy impacts the thermoelectric properties of disordered organic semiconductors and how this depends on the morphology the polymer crystallizes in. The model rationalizes experiments on PBTTT and it is examined how parameters affecting the length scales and the structural order of the system impact the power factor.

\noindent
\textbf{Keyword:} organic thermoelectrics

\noindent
\textbf{Authors:} Dorothea Scheunemann, Vishnu Vijayakumar, Huiyan Zeng, Pablo Durand, Nicolas Leclerc, Martin Brinkmann, and Martijn Kemerink

\noindent
\textbf{Title:} Rubbing and Drawing: Generic Ways to Improve the Thermoelectric Power Factor of Organic Semiconductors?

\begin{figure}
\centering
\includegraphics{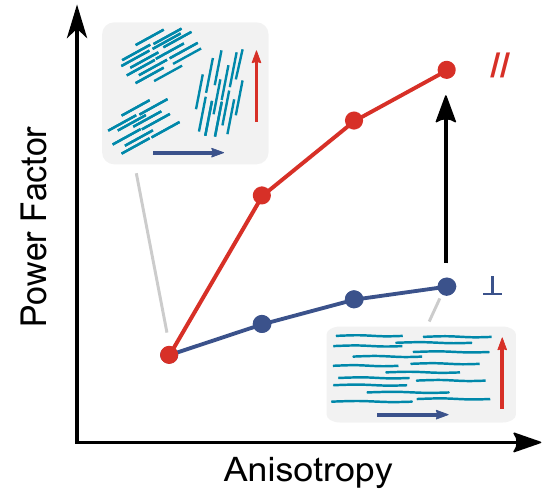}%
\label{fig:TOC}
\end{figure}

\end{document}